\documentclass[11pt]{article}

\usepackage{amssymb}
\usepackage[pdftex]{color,graphicx}
\usepackage[paperwidth=5.3in, paperheight=7.5in, top=.55in, bottom=.5in, left=.14in, right=.14in]{geometry}
\usepackage{textcomp}

\newcommand{\ipar}{\begin{list}{\hbox{$\:$}}{} \item[] }
\newcommand{\rapi}{\end{list}}

\begin{document}

\title{\bf Faster Maximium Priority Matchings \\in Bipartite Graphs}
  
\author{Jonathan Turner}
\date{\normalsize {\sc wucse}-2015-08}
\maketitle

\begin{abstract}
A maximum priority matching is a matching in an undirected graph that maximizes
a {\sl priority score} defined with respect to given vertex priorities.
An earlier paper showed how to find maximum priority matchings in unweighted graphs.
This paper describes an algorithm for bipartite graphs that is faster when the
number of distinct priority classes is limited.
For graphs with $k$ distinct priority classes it runs in $O(kmn^{1/2})$ time,
where $n$ is the number of vertices in the graph and $m$ is the number of edges.
\end{abstract}

\newpage

\pagestyle{plain}

The maximum priority matching problem was introduced in~\cite{turner-2015}.
In this problem, each vertex has an integer-valued {\sl priority} and the objective
is to find a matching that maximizes a {\sl priority score} defined with respect to
these values. The priority score is defined as the $n$-ary number in which
the $i$-th most-significant digit is the number of matched vertices with priority $i$.
The earlier paper described an algorithm for finding maximum priority matchings in $O(mn)$ time
that is based on Edmonds' algorithm for the maximum size matching problem~\cite{ed65,ga76,mv80}.

In a private communication, Tarjan observed that the 2-priority case for bipartite graphs
could be solved in $O(mn^{1/2})$ time using a recent algorithm for weighted
matchings in bipartite graphs~\cite{duan-2011,goldberg-2015,ramshaw-2012}.
One assigns a weight of 0, 1 or 2 to each edge based on the number of high priority vertices it is
incident to. A maximum weight matching of this graph matches the largest possible number
of high priority vertices. It can then be extended to a maximum priority matching using
an algorithm by Hopcroft and Karp~\cite{hk73}. 
Tarjan speculated that this approach might be extended to
handle multiple priority classes. This paper takes a different approach, yielding a simpler
algorithm that handles the $k$-priority case in $O(kmn^{1/2})$ time,
making it faster than the earlier algorithm when $k$ grows more slowly than $n^{1/2}$.

We assume that the reader is familiar with~\cite{turner-2015}, which describes
a generalization of the augmenting path algorithm of Edmonds.
In this algorithm there are two types of augmenting paths.
Odd-length augmenting paths increase the number of edges in the current matching
and may also increase its priority score, while even-length augmenting paths
increase the priority score without changing the size of the matching.

Here, we take a slightly different approach.
Let $G=(V,E)$ be an undirected bipartite graph with vertex priorities $p(u)$
and let $V_1$ and $V_2$ define the bipartition of $G$ (that is, all edges join
vertices in $V_1$ to vertices in $V_2$).
The algorithm starts by finding a maximum size matching,
then sets an integer variable $i$ to the index of the first non-empty priority class 
and repeats the following step until all non-empty priority classes have been processed.
\ipar
While there are even-length $i$-augmenting paths with an unmatched vertex in $V_1$,
find such a path and reverse the matching status of its edges.
While there are even-length $i$-augmenting paths with an unmatched vertex in $V_2$,
find such a path and reverse the matching status of its edges.
Advance $i$ to the index of the next non-empty priority class.
\rapi
The augmenting path searches in each step can be implemented by solving
a pair of maximum flow problems on {\sl unit graphs}. These can be solved in $O(mn^{1/2})$
time using Dinic's algorithm~\cite{tarjan-1983}. The original maximum size matching can
also be found in $O(mn^{1/2})$ time using the Hopcroft-Karp algorithm,
yielding an overall time bound of $O(kmn^{1/2})$ for $k$ priority classes.

Let $M_1$ be the matching at the start of step $i$. We construct an instance of the maximum
flow problem $X_1=(W_1,F_1)$ as follows.
\begin{eqnarray*}
W_1&=& V \cup \{s,t\} \\
F_1 &=& \{(u,v) \;|\; \hbox{$u\in V_1$, $v\in V_2$ and $\{u,v\}\not\in M_1$}\} \\
&\cup& \{(v,u) \;|\; \hbox{$u\in V_1$, $v\in V_2$ and $\{u,v\}\in M_1$}\} \\
&\cup& \{(s,u) \;|\; \hbox{$u\in V_1$ is unmatched and $p(u)=i$}\} \\
&\cup& \{(u,t) \;|\; \hbox{$u\in V_1$ is matched and $p(u)>i$}\}
\end{eqnarray*}
All edges are assigned a capacity of 1. Observe that $X_1$ is a unit graph, since each
vertex has at most one incoming or one outgoing edge. A maximum flow $f_1$ on $X_1$ can
be decomposed into a set of augmenting paths in $X_1$. Each of these paths
corresponds directly to an $i$-augmenting path in $G$ and consequently, $f_1$
defines a modified matching $M_2$.
\begin{eqnarray*}
M_2 &=& \{\{u,v\} \;|\; \hbox{$u\in V_1$, $v\in V_2$, $(u,v)\in F_1$ and $f_1(u,v)=1$}\} \\
&\cup& \{\{u,v\} \;|\; \hbox{$u\in V_1$, $v\in V_2$, $(v,u)\in F_1$ and $f_1(v,u)=0$}\} 
\end{eqnarray*}
Note that every vertex with priority $\leq i$ that is matched in $M_1$ is also matched in $M_2$.

Next, the algorithm constructs a second flow problem $X_2=(W_2,F_2)$.
\begin{eqnarray*}
W_2&=& V \cup \{s,t\} \\
F_2 &=& \{(u,v) \;|\; \hbox{$u\in V_1$, $v\in V_2$ and $\{u,v\}\not\in M_2$}\} \\
&\cup& \{(v,u) \;|\; \hbox{$u\in V_1$, $v\in V_2$ and $\{u,v\}\in M_2$}\} \\
&\cup& \{(s,v) \;|\; \hbox{$v\in V_2$ is matched and $p(u)>i$}\} \\
&\cup& \{(v,t) \;|\; \hbox{$v\in V_2$ is unmatched and $p(u)=i$}\}
\end{eqnarray*}
All edges are again assigned a capacity of 1. A maximum flow $f_2$ on $X_2$ 
defines a modified matching $M_3$.
\begin{eqnarray*}
M_3 &=& \{\{u,v\} \;|\; \hbox{$u\in V_1$, $v\in V_2$, $(u,v)\in F_2$  and $f_2(u,v)=1$}\} \\
&& \cup \{\{u,v\} \;|\; \hbox{$u\in V_1$, $v\in V_2$, $(v,u)\in F_2$ and $f_2(v,u)=0$}\} 
\end{eqnarray*}
Observe that every vertex with priority $\leq i$ that is matched in $M_1$ 
is also matched in $M_3$.


\begin{thebibliography}{99}

\bibitem{duan-2011}
Duan, Ran, Seth Pettie, and Hsin-Hao Su. 
``Scaling algorithms for approximate and exact maximum weight matching,'' 
{\sl Computing Research Repository} (CoRR), abs/1112.0790, 2011.

\bibitem{ed65}
Edmonds, Jack. ``Paths, trees and flowers,'' {\sl Canadian Journal of Mathematics}, 
1965, pp. 449--467.

\bibitem{ga76}
Gabow, Harold N.
``An efficient implementation of Edmonds' algorithm for maximum matching on graphs,''
{\sl Journal of the Association for Computing Machinery}, 1976, pp. 221--234.

\bibitem{goldberg-2015}
Goldberg, Andrew V., Haim Kaplan, Sagi Hed, and Robert E. Tarjan.
``Minimum cost flows in graphs with unit capacities,''
{\sl Symposium on Theoretical Aspects of Computer Science} (STACS), 2015.

\bibitem{hk73}
Hopcroft, John E. and Richard M. Karp.
``An $O(n^{5/2}$ algorithm for maximum matching in bipartite graphs,''
{\sl SIAM Journal on Computing}, 1973, pp 225--231.

\bibitem{mv80}
Micali, Silvio. and V. V. Vazirani.
``An $O(\sqrt{|V|}\cdot |E|)$ algorithm for finding maximum matchings in general graphs,''
{\sl IEEE Symposium on the Foundations of Computer Science (FOCS)}, 1980, pp. 17-27.

\bibitem{ramshaw-2012}
Ramshaw, L., and Robert E. Tarjan.
``A weight-scaling algorithm for min-cost imperfect matchings in bipartite graphs". 
In {\sl IEEE Symposium on the Foundations of Computer Science}, pages 581–590, 2012.

\bibitem{tarjan-1983}
Tarjan, Robert E.
{\sl Data structures and network algorithms}.
Society for Industrial and Applied Mathematics, 1983.

\bibitem{turner-2015}
Turner, Jonathan S.
``Maximum priority matchings,''
Washington University Computer Science and Engineering Department technical report,
{\sc wucs-2015-06}, 2015.
\end{thebibliography}
\end{document}